\documentclass[review, 3p, times]{elsarticle}

\usepackage{lineno}       
\usepackage{hyperref}     
\usepackage{graphicx}     
\usepackage{amsmath}      
\usepackage{booktabs}     
\usepackage{filecontents} 
\usepackage{tikz}

\usetikzlibrary{arrows.meta}
\usetikzlibrary{positioning}
\usetikzlibrary{shadows}
\usetikzlibrary{calc}
\usetikzlibrary{shadows.blur}

\journal{}

\begin{document}

\begin{frontmatter}


\title{UniCrop: A Universal, Multi-Source Data Engineering Pipeline for Scalable Crop Yield Prediction}

\author[1]{Emiliya Khidirova}
\address[1]{Cardiff University, School of Mathematics, CF24 4AG UK.}
\ead{emiliyakhidirova@gmail.com}
\author[2]{Oktay Karakuş\corref{cor1}}
\address[2]{Cardiff University, School of Computer Science \& Informatics, CF24 4AG UK}
\ead{karakuso@cardiff.ac.uk}
\cortext[cor1]{Corresponding author}

\begin{abstract}
Accurate crop yield prediction increasingly relies on diverse data streams, including satellite observations, meteorological reanalysis, soil composition, and topographic information. However, despite rapid advances in machine learning, most existing approaches remain crop- or region-specific and require substantial bespoke data engineering efforts. This limits scalability, reproducibility, and operational deployment. 
This study introduces \textbf{UniCrop}, a universal and reusable data pipeline designed to automate the acquisition, cleaning, harmonisation, and feature engineering of multi-source environmental data for crop yield prediction. For any given location, crop type, and temporal window, UniCrop automatically retrieves, harmonises, and engineers over 200 environmental variables from heterogeneous satellite, climate, soil, and topographic sources (Sentinel-1/2, MODIS, ERA5-Land, NASA POWER, SoilGrids, and SRTM), reducing them to a compact, analysis-ready feature set utilising a structured feature reduction workflow with minimum redundancy maximum relevance (mRMR). 
To validate the pipeline, UniCrop was applied to a rice yield dataset comprising 557 field observations. Using only the selected 15 features, four baseline machine-learning models (LightGBM, Random Forest, Support Vector Regression, and ElasticNet) were trained using rigorous cross-validation. LightGBM achieved the best single-model performance (RMSE = 465.1\,kg/ha, $R^{2} = 0.6576$), while a constrained ensemble of all baselines further improved accuracy (RMSE = 463.2\,kg/ha, $R^{2} = 0.6604$). SHAP analysis confirmed agronomically plausible relationships and demonstrated how UniCrop leverages multi-modal predictors. 
UniCrop contributes a scalable and transparent data-engineering framework that addresses the primary bottleneck in operational crop yield modelling: the preparation of consistent and harmonised multi-source data. By decoupling data specification from implementation and supporting any crop, region, and time frame through simple configuration updates, UniCrop provides a practical foundation for transferable, high-quality agricultural analytics at scale. The code and implementation documentation are shared in \url{https://github.com/CoDIS-Lab/UniCrop}.
\end{abstract}

\begin{keyword} 
Crop yield prediction \sep remote sensing \sep data harmonisation \sep agricultural machine learning \sep feature engineering \sep multi-source integration \sep Sentinel \sep ERA5 \sep SoilGrids \sep data pipelines.
\end{keyword}

\end{frontmatter}


\section{Introduction}

Agriculture plays a central role in global food security and economic stability, yet it faces intensifying pressures from climate change, population growth, and increasingly volatile weather patterns \citep{Jabed2024, Li2025}. Accurate crop yield prediction is essential for informing government policies, stabilising supply chains, guiding agribusiness logistics, and supporting farmers' decisions on irrigation, fertiliser application, and harvest scheduling. These needs have grown more urgent as production shocks in one region can propagate rapidly across international markets \citep{Morales2023}.

Significant advances in Earth Observation (EO), agro-meteorology, and machine learning (ML) have expanded the potential for data-driven agricultural forecasting. Open EO programmes such as the Copernicus Sentinel missions provide optical, radar, and atmospheric measurements at high spatial and temporal resolution, enabling detailed monitoring of crop conditions \citep{He2025, Nitu2025}. Long-term vegetation indices from MODIS facilitate phenological analysis across large regions \citep{MdTahir2024}. Climate datasets such as ERA5-Land and NASA POWER offer globally consistent information on temperature, precipitation, radiation, humidity, and wind, key drivers of agricultural productivity \citep{Wang2024}. Complementary environmental datasets from SoilGrids and SRTM further describe soil texture, carbon content, pH, elevation, slope, and microclimatic influences \citep{Geng2025}.

Despite these advances, the practical development of yield prediction models remains hindered by a persistent data-engineering bottleneck. Most studies construct bespoke pipelines tailored to a specific crop, region, or time frame, requiring substantial manual effort to integrate heterogeneous datasets and harmonise spatial and temporal resolutions \citep{Jabed2024}. Even state-of-the-art multimodal deep learning approaches, such as recently proposed ensemble and fusion systems for rice yield prediction \citep{Yewle2025}, rely on complex, labour-intensive preprocessing workflows. As noted in recent reviews, the central challenge is increasingly one of scalable and reproducible data preparation rather than modelling innovation \citep{Morales2023}.

To address this challenge, we introduce \textbf{UniCrop}, a universal, configuration-driven data pipeline that automates the acquisition, harmonisation, and transformation of multi-source environmental data for crop yield prediction. UniCrop separates the specification of required variables from implementation, enabling users to adapt the pipeline to new crops or regions by modifying a simple configuration file. The system integrates optical and radar EO data, vegetation indices, climate reanalysis, agro-climatological variables, soil composition, and topographic layers into a unified, analysis-ready dataset. 

We validate UniCrop using a real-world rice yield dataset comprising 557 field observations. The case study demonstrates that high-quality and consistent multi-source features generated by UniCrop support accurate predictions using standard machine-learning models, while interpretability analysis confirms that the pipeline captures agronomically meaningful relationships. Together, these contributions establish UniCrop as a robust and scalable foundation for multi-crop, multi-region yield prediction.

UniCrop is not intended to replace crop-specific modelling expertise or advanced learning architectures. Instead, it aims to provide a robust, reusable, and transparent data foundation upon which a wide range of statistical and machine-learning models can be applied consistently across crops and regions.

\section{Background}

Crop yield prediction has evolved substantially over the last two decades, driven by advances in remote sensing, agro-meteorological modelling, and machine learning (ML). Early approaches predominantly relied on statistical regressions and empirical models calibrated for individual crops and regions. Recent studies, however, increasingly integrate multi-source environmental data with ML architectures to capture non-linear crop–environment interactions. Despite strong progress, persistent challenges remain in scalability, data heterogeneity, and operational generalisation. This section reviews the key developments in three interconnected domains: (i) satellite-based remote sensing, (ii) environmental and agro-climatic data integration, and (iii) machine-learning approaches for yield prediction. We conclude by identifying the unresolved limitations in current practice that motivate the UniCrop framework.

\subsection{Satellite-Based Optical and Radar Remote Sensing}

Remote sensing has become central to agricultural monitoring, supporting vegetation assessment, phenology tracking, flood mapping, and yield forecasting. Optical sensors, particularly the Multispectral Instrument aboard Sentinel-2, deliver high-resolution reflectance data used to derive vegetation indices such as the Normalised Difference Vegetation Index (NDVI) and Enhanced Vegetation Index (EVI). These indices provide proxies for canopy greenness, biomass accumulation, and stress responses \citep{Nitu2025,He2025}. However, optical data suffer from cloud contamination, especially in monsoon-affected or tropical regions where agricultural monitoring is most needed.

Synthetic Aperture Radar (SAR) addresses this limitation by providing cloud-penetrating microwave observations. Sentinel-1 SAR data capture structural and moisture-related properties of crop canopies, enabling robust monitoring under all-weather conditions \citep{Keerthana2022}. Studies increasingly fuse optical and SAR observations to exploit their complementary characteristics. Recent transformer-based fusion models demonstrate performance gains for rice mapping and phenology recognition \citep{He2025}, while multimodal satellite–weather systems have been shown to improve rice yield prediction accuracy through ensemble deep learning architectures \citep{Yewle2025}. These trends highlight the power of EO fusion but also underscore the data-engineering workload inherent in multi-sensor integration.

\subsection{Coarse-Resolution Time Series and Vegetation Dynamics}

Beyond high-resolution imagery, coarse-resolution products such as MODIS remain essential due to their high temporal frequency and global coverage. MODIS vegetation indices (MOD13), leaf area index (MOD15), evapotranspiration (MOD16), and gross primary productivity (MOD17) have been widely used to monitor seasonal dynamics of crop development \citep{MdTahir2024}. Their consistent, cloud-screened composites enable long-term trend analysis and facilitate large-scale or multi-region yield modelling. Time-series modelling using NDVI curves, phenological metrics, or harmonised Landsat–MODIS series has been applied to wheat, maize, rice, soybean, and potato systems worldwide \citep{Alsaber2025,Liang2024}. However, harmonising MODIS with higher-resolution datasets requires careful temporal alignment and spatial reduction, which is often implemented manually.

\subsection{Integrating Climatic, Soil, and Topographic Drivers}

Crop development is strongly influenced by weather variability and environmental conditions. Climate reanalysis products such as ERA5-Land provide consistent global records of temperature, radiation, precipitation, and soil moisture, critical for modelling crop–climate interactions \citep{Li2025}. Similarly, NASA POWER offers agro-climatology-ready variables including vapour pressure deficit (VPD), dewpoint temperature, wind speed, and diurnal temperature range (DTR). These variables capture key stressors associated with heatwaves, humidity fluctuations, and atmospheric dryness.

Static environmental properties further influence crop growth. SoilGrids provides global soil texture, carbon, pH, and bulk density maps at 250\,m resolution, enabling soil–water and nutrient availability modelling \citep{Geng2025}. Terrain data from the Shuttle Radar Topography Mission (SRTM) inform hydrological behaviour, slope-driven erosion, and microclimate \citep{SRTM}. Integrating these diverse datasets requires careful standardisation, reprojection, temporal matching, and provenance tracking, steps that are rarely automated in existing agricultural analytics workflows.

\subsection{Machine Learning Approaches for Yield Prediction}

Machine learning has become a dominant tool in data-driven yield prediction, offering the ability to model complex, non-linear relationships across diverse input features. Ensemble tree-based models such as Random Forests, Gradient Boosting Machines, and LightGBM frequently outperform classical regressions due to their robustness and ability to handle noisy inputs \citep{Jabed2024,Morales2023}. Recent studies have also explored neural architectures, including convolutional neural networks (CNNs) for spatial imagery and recurrent or LSTM networks for time-series climate and phenology modelling \citep{Joshi2025}. Deep ensembles combining EO imagery, meteorology, and crop-growth indicators have demonstrated strong performance for rice, maize, and wheat yield prediction \citep{Yewle2025,Das2023}.

Despite these modelling innovations, several challenges persist. Deep learning architectures require large datasets and are sensitive to missing or misaligned inputs. More importantly, many state-of-the-art models perform only marginally better than simpler baselines when trained on datasets that are inconsistently engineered or incomplete \citep{Morales2023}. This underscores a critical insight: data quality and harmonisation often determine predictive performance more than model complexity.

\subsection{Limitations of Current Pipelines and Motivation for UniCrop}

Across the literature, most studies develop bespoke workflows tailored to a single crop, region, or satellite product. These workflows are often complex, manually curated, and difficult to reproduce. The lack of standardised pipelines leads to siloed methodologies, limited transferability, and substantial entry barriers for agricultural researchers. As noted in several reviews \citep{Jabed2024,Wang2024}, the primary obstacle to scalable yield prediction is not model design but the engineering required to gather, harmonise, and prepare multi-source environmental data.

UniCrop directly addresses these limitations by providing a universal, configuration-driven, multi-source data pipeline that automates acquisition, cleaning, temporal alignment, spatial aggregation, feature engineering, and selection across EO, climate, soil, and topographic datasets. By standardising the data-engineering foundation, UniCrop enables reproducible and transferable yield modelling regardless of crop type or geographic context.

\section{Materials and Methods}

UniCrop is designed as a universal, reusable, and configuration-driven data pipeline that automates the process of acquiring, cleaning, harmonising, and transforming multi-source environmental data into an analysis-ready dataset for crop yield prediction. The system decouples the choice of data sources from the underlying implementation, enabling rapid portability across crops, regions, and temporal windows. This section details the full UniCrop methodology, from pipeline design and data ingestion to feature engineering, selection, and model training.

\subsection{High-Level Pipeline Architecture}
\begin{figure}[ht!]
\centering
\begin{tikzpicture}[
    scale=0.7,
    every node/.style={font=\normalsize, transform shape},
    node distance=0.6cm and 1cm,
    process/.style={rectangle, rounded corners, draw=black, very thick,
                     text centered, align=center,
                     minimum height=1.5cm, minimum width=3.5cm,
                     fill=gray!10},
    datasource/.style={rectangle, rounded corners, draw=black, thick,
                       text centered, align=center,
                       minimum height=0.9cm, minimum width=2.5cm,
                       fill=blue!10},
    arrow/.style={-{Latex[length=3mm,width=2mm]}, thick},
]

\node[datasource] (s1) {Sentinel-1\\(SAR)};
\node[datasource, right=of s1] (s2) {Sentinel-2\\(Optical)};
\node[datasource, right=of s2] (modis) {MODIS\\(ET, NDVI, fPAR)};
\node[datasource, right=of modis] (era5) {ERA5-Land};
\node[datasource, right=of era5] (power) {NASA POWER};
\node[datasource, right=of power] (soil) {SoilGrids};
\node[datasource, right=of soil] (srtm) {SRTM\\(Topography)};

\node[process, below=1.8cm of s1] (acq) {Multi-source\\Data Acquisition};

\foreach \src in {s2,modis,era5,power,soil,srtm} {
    \draw[arrow] (\src.south) -- ++(0,-0.3) |- (acq);
}
\draw[arrow] (s1.south) -- (acq.north);

\node[process, below=1.8cm of acq] (harm) {Data Harmonisation \&\\Temporal/Spatial Alignment};

\draw[arrow] (acq) -- (harm);

\node[process, right=1.8cm of harm] (fe) {Feature Engineering};

\draw[arrow] (harm) -- (fe);

\node[process, right=1.6cm of fe] (fs) {mRMR Feature Selection\\(Compact 15 Feature Set)};

\draw[arrow] (fe) -- (fs);

\node[process, right=1.6cm of fs] (mod) {Baseline ML Modelling\\(LightGBM, RF, SVR, ElasticNet)};

\draw[arrow] (fs) -- (mod);

\node[process, below=1.6cm of mod] (eval) {Ensemble \& Interpretability\\(SHAP, Metrics)};

\draw[arrow] (mod.east) -- ++(0.5,-0.5) |- (eval.east);

\end{tikzpicture}
\caption{Hybrid schematic of the UniCrop pipeline. Multiple environmental data sources are ingested in parallel, harmonised into a unified structure, enriched with agronomic features, reduced via mRMR into a compact predictor subset, and finally used for baseline modelling and ensemble evaluation.}
\label{fig:pipeline}
\end{figure}
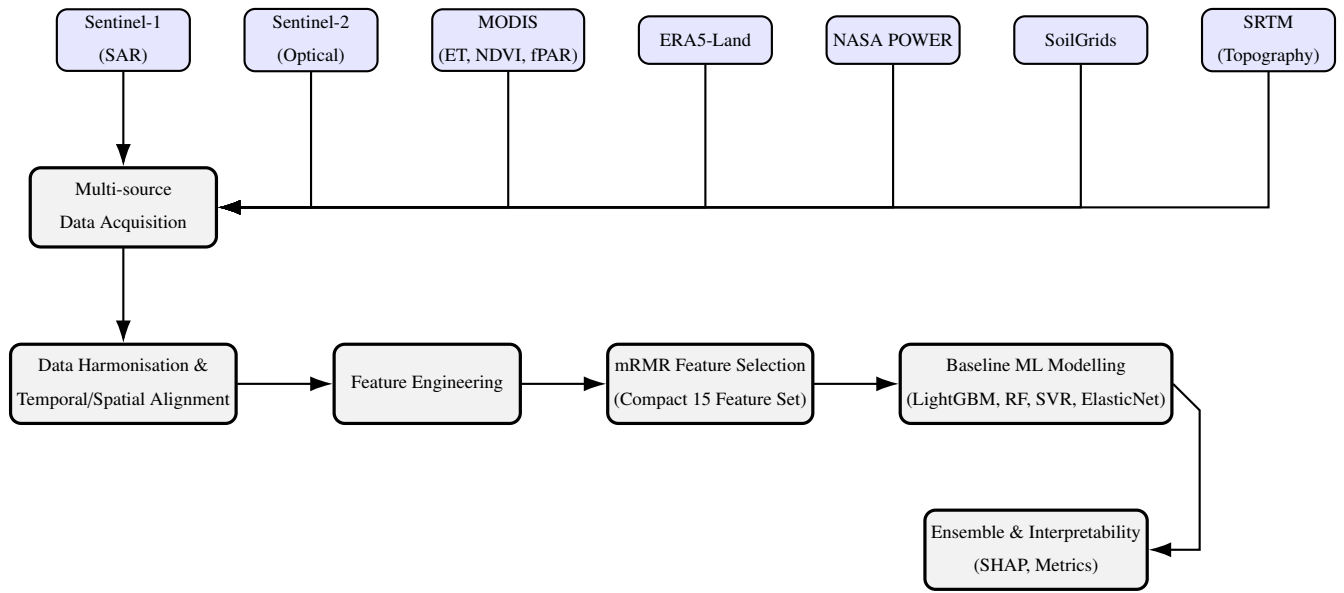

The UniCrop architecture (Figure~\ref{fig:pipeline}) is structured into five modular stages:

\begin{enumerate}
    \item \textbf{Configuration and Planning}: Users specify required features through a structured feature mapping file and provide field-level observations (latitude, longitude, dates). These inputs drive a dynamically constructed \emph{fetch plan}.
    \item \textbf{Data Acquisition}: Multi-source environmental data are collected from satellite, climate, soil, and topographic repositories through Google Earth Engine (GEE) and public APIs.
    \item \textbf{Data Harmonisation}: All source outputs are temporally and spatially aligned, cleaned, standardised, and merged into a unified table with complete provenance records.
    \item \textbf{Feature Engineering and Selection}: Statistical screening, agronomic feature engineering, and minimum redundancy maximum relevance (mRMR) selection are applied to derive a compact subset of informative predictors.
    \item \textbf{Baseline Model Training and Evaluation}: Selected features are used to train multiple baseline models under rigorous cross-validation, followed by a constrained ensemble.
\end{enumerate}

This modular design ensures that extending UniCrop to new crops, regions, or temporal resolutions requires changes only to the configuration layer, without modification of the underlying pipeline logic.

\subsection{Feature Mapping and Fetch Plan Generation}

\begin{figure}[ht!]
\centering
\begin{tikzpicture}[
    scale=0.80,
    every node/.style={font=\small, transform shape},
    table/.style={rectangle, draw=black, thick, rounded corners,
                  fill=blue!10, minimum width=12.5cm, minimum height=4.2cm,
                  align=center},
    process/.style={rectangle, rounded corners, draw=black, thick,
                    minimum width=4.4cm, minimum height=1.3cm,
                    align=center, fill=gray!10},
    arrow/.style={-{Latex[length=3mm,width=2mm]}, thick}
]

\node[table] (mapping) {
\begin{tabular}{l l l l p{5.8cm}}
\textbf{Key Variable} & \textbf{API Param} & \textbf{Source Dataset} & \textbf{Platform} & \textbf{Notes / Derivation} \\
\hline
Aspect & aspect & SRTM (DEM) & USGS/SRTMGL1\_003 & Derived from DEM using \texttt{terrain()} function in GEE. \\
ET (Evapotranspiration) & ET & MOD16A2 (MODIS) & MODIS/006/MOD16A2 & Daily evapotranspiration (mm/day). \\
EVI & Derived & Sentinel-2 & COPERNICUS/S2 & EVI $= 2.5\frac{NIR - RED}{NIR + 6\,RED - 7.5\,BLUE + 1}$. \\
Elevation & elevation & SRTM (DEM) & USGS/SRTMGL1\_003 & Elevation (m) from SRTM DEM. \\
Irrigation & Derived & ERA5-Land & ECMWF/ERA5\_LAND/DAILY & Irrigation = max(0, PEV – TP). \\
Total Precipitation & total\_precipitation & ERA5-Land & ECMWF/ERA5\_LAND/DAILY & Accumulated precipitation over time step. \\
\end{tabular}
\\[4pt]
\textit{Excerpt from \texttt{unicrop\_feature\_mapping.csv}}
};

\node[process, below=1.5cm of mapping] (fetchplan)
{Fetch Plan\\(Fields × Variables)};

\draw[arrow] (mapping.south) -- (fetchplan.north);

\node[process, right=3.2cm of fetchplan] (acq)
{Automated Data\\Acquisition};

\draw[arrow] (fetchplan.east) -- (acq.west);

\end{tikzpicture}
\caption{Schematic representation of UniCrop’s feature mapping system, illustrated with real examples from the mapping file. 
The mapping table defines variables, API parameters, datasets, and derivations used to build the automated fetch plan.}
\label{fig:featuremapping}
\end{figure}
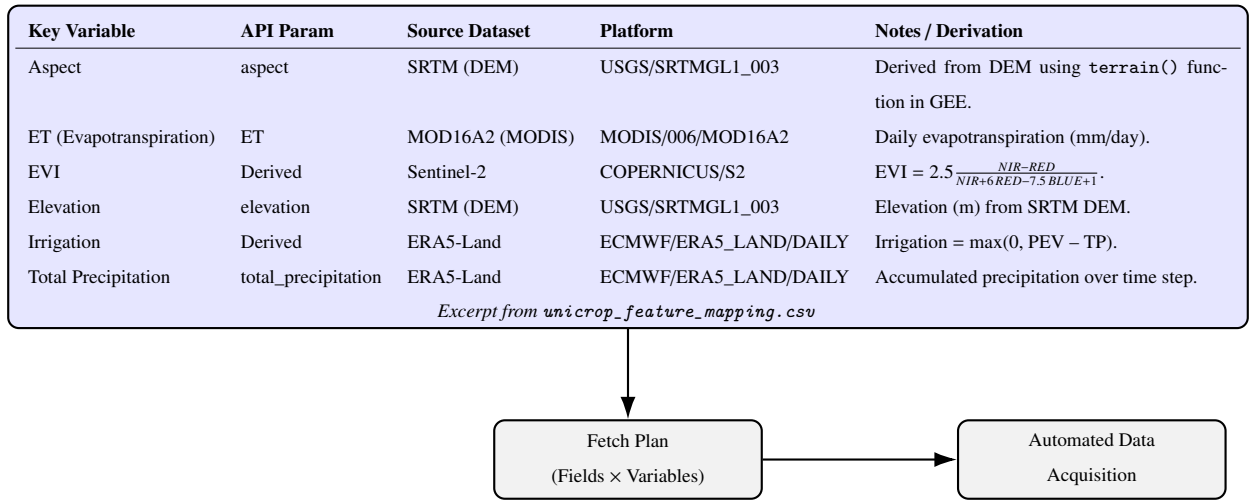

The feature mapping file serves as the central configuration layer that defines all environmental variables required by UniCrop. Each entry specifies the key variable name, the API parameter used for retrieval, the source dataset, the platform (e.g.\ GEE, NASA POWER, SoilGrids), and any associated derivation or calculation rules. This structured mapping enables UniCrop to separate data specification from data implementation, ensuring that users can adapt the pipeline to new crops or regions simply by modifying a human-readable configuration file rather than altering code. An illustrative excerpt is shown in Figure~\ref{fig:featuremapping}, which highlights how variables such as EVI, evapotranspiration, elevation, and total precipitation are declaratively defined. During execution, UniCrop expands this mapping into a comprehensive \emph{fetch plan} that enumerates all field--variable combinations, forming the basis for automated multi-source data acquisition.

\subsubsection{Feature Mapping Specification}

All features required by the user are defined in a single CSV specification, \texttt{unicrop\_feature\_mapping.csv}. Each row corresponds to a data variable and includes:

\begin{itemize}
    \item \textbf{Key Variable}: Human-readable identifier (e.g., \texttt{NDVI}, \texttt{T2M}, \texttt{soil\_carbon}).
    \item \textbf{API Parameter}: Data-source-specific token controlling retrieval (e.g., Sentinel-2 band names or NASA POWER API codes).
    \item \textbf{Source Dataset}: The product identifier (e.g., ERA5-Land, MODIS MOD13Q1, Copernicus Sentinel-1).
    \item \textbf{Platform}: Retrieval route (e.g., GEE collection ID, API endpoint).
    \item \textbf{Notes/Derivation}: Optional derivation instructions, including formulas for indices.
\end{itemize}

By separating data specification from implementation, the feature mapping mechanism also improves reproducibility, as the exact data requirements for a study can be shared, audited, and reused independently of code.

\subsubsection{Fetch Plan Construction}

The system automatically generates a \emph{fetch plan} that enumerates all required retrieval operations. This is constructed by:

\begin{enumerate}
    \item Cleaning and standardising the input field table (coordinates, dates, identifiers);
    \item Normalising the feature mapping (column names, dataset identifiers, API parameters);
    \item Creating a Cartesian product between input records and feature requests.
\end{enumerate}

The resulting \texttt{fetch\_plan.csv} defines, for each field-date record, which variable to retrieve, from which source, and under what parameters. Duplicates, invalid coordinates, and malformed rows are automatically removed.

\subsection{Data Sources and Acquisition Methods}

UniCrop integrates six major data categories: optical satellite imagery, radar backscatter, coarse-resolution vegetation products, meteorological reanalysis, agro-climatological data, and environmental context (soil and topography). Table~\ref{tab:datasources} summarises all sources.

\begin{table}[h!]
\centering
\caption{Summary of data sources integrated in UniCrop.}
\label{tab:datasources}
\begin{tabular}{p{3.95cm} p{2.4cm} p{7.8cm}}
\toprule
\textbf{Source} & \textbf{Type} & \textbf{Data Provided} \\
\midrule
Sentinel-2 MSI & Optical RS & Reflectance bands; NDVI, EVI, SAVI, NDRE, CI\textsubscript{red-edge}; canopy biophysical indices \\
Sentinel-1 SAR & Radar RS & VV, VH backscatter; VV/VH ratio; Radar Vegetation Index (RVI); texture features \\
MODIS (MOD13, MOD15, MOD16, MOD17) & Vegetation & NDVI/EVI (16-day), fPAR, LAI, ET, GPP \\
ERA5-Land & Climate & Temperature, precipitation, soil moisture, radiation, evaporation (hourly) \\
NASA POWER & Agro-climate & T2M, T2M\_MIN/MAX, RH2M, VPD, DTR, wind speed, solar radiation \\
SoilGrids & Soil & Soil texture, SOC, bulk density, nitrogen, pH (0–30\,cm) \\
SRTM & Topography & Elevation, slope, aspect, terrain derivatives \\
\bottomrule
\end{tabular}
\end{table}

\subsection{Data Harmonisation and Master Table Construction}

All source-specific outputs are merged into a master time-series table keyed by latitude, longitude, and date. UniCrop applies:

\begin{itemize}
    \item ISO-standard date formatting,
    \item numeric casting and float precision enforcement,
    \item duplicate removal with prioritisation of data completeness,
    \item suffix restoration for consistent naming (e.g., elevation, slope),
    \item generation of \texttt{unicrop\_columns\_manifest.csv} for provenance.
\end{itemize}

Missingness is preserved until imputation within cross-validation to prevent leakage.

\subsection{Feature Engineering}

UniCrop generates additional features capturing agronomic processes:

\begin{itemize}
    \item \textbf{Growing Degree Days (GDD)}:
    \[
    \mathrm{GDD}_{\mathrm{base10}} = \max\left(0, \frac{T_{\mathrm{max}} + T_{\mathrm{min}}}{2} - 10\right)
    \]
    \item \textbf{Chill nights}: count of days where $T_{\mathrm{min}} < 15^\circ$C.
    \item \textbf{Vegetation dynamics}: seasonal amplitude of NDVI/EVI.
    \item \textbf{SAR texture metrics}: variability in VV/VH backscatter.
    \item \textbf{Soil--climate interactions}: e.g., clay $\times$ radiation, elevation $\times$ temperature.
\end{itemize}

All engineered features undergo type validation and schema alignment, and they are designed to be crop-agnostic and derived solely from environmental signals, ensuring applicability across different cropping systems without requiring crop-specific calibration.

\subsection{Statistical Screening and mRMR Feature Selection}

Feature screening proceeds in several stages designed to remove redundant or uninformative predictors before applying a more principled selection method. First, features with near-zero variance are discarded. Second, highly collinear variables ($r \ge 0.98$) are pruned by retaining only the feature with the highest bivariate association with yield as measured by mutual information. Third, a family-preservation heuristic ensures that at least one feature from each major environmental data family (meteorology, vegetation, SAR, soil, topography) is retained in the candidate pool, preventing over-selection from any dominant modality.

The final selection step employs the minimum redundancy maximum relevance (mRMR) algorithm, a widely used filter-based technique for identifying compact feature subsets. Let $S$ denote the selected feature set and $f$ a candidate feature. The \emph{relevance} of $f$ to the target variable $y$ is quantified using mutual information,
\[
I(f; y) = \iint p(f, y) \log \frac{p(f, y)}{p(f)\,p(y)} \, df\, dy,
\]
which captures general (non-linear) dependencies beyond simple correlation. In practice, an ensemble relevance score combining mutual information, Pearson correlation, and Spearman rank correlation is used to enhance robustness across varying feature types.

The \emph{redundancy} between a candidate feature $f$ and an existing selected feature $s \in S$ is computed as
\[
R(f; S) = \frac{1}{|S|} \sum_{s \in S} I(f; s),
\]
penalising features that provide overlapping information. The mRMR objective seeks to maximise the difference between relevance and redundancy:
\[
\max_{f \notin S} \left[ I(f; y) - R(f; S) \right],
\]
or equivalently, maximise the relevance-to-redundancy ratio,
\[
\max_{f \notin S} \left[ \frac{I(f; y)}{R(f; S) + \epsilon} \right],
\]
where $\epsilon$ is a small constant preventing division by zero.

Features are added iteratively until the desired subset size is reached. In this study, a compact set of 15 predictors (UniCrop default value is 15, and can be set by users to any value) is selected \emph{independently within each cross-validation fold} to ensure that no information from the test partitions influences feature selection, thereby preventing leakage. The resulting subset preserves diversity across environmental data families while maintaining a strong aggregate relevance to rice yield.

The resulting compact feature set prioritises interpretability and generalisability over marginal performance gains, aligning to produce robust baseline models rather than maximally tuned predictors.

\subsection{Model Training and Evaluation}

\subsubsection{Cross-Validation Strategy}

Model evaluation is conducted using a 5-fold shuffled cross-validation scheme at the field level. To ensure an unbiased assessment, all preprocessing steps, including imputation, scaling, statistical screening, feature engineering, and mRMR feature selection, are performed \emph{independently within each training fold}. Validation folds remain completely unseen during preprocessing, preventing information leakage and yielding a realistic estimate of out-of-sample performance.

\subsubsection{Imputation and Scaling}

Missing values and scale heterogeneity are addressed using a family-aware preprocessing strategy designed to preserve physical meaning while preventing information leakage. All imputation and scaling steps are performed \emph{within each cross-validation fold}.

\begin{itemize}
    \item \textbf{Meteorological variables}: multivariate iterative imputation is applied to exploit cross-variable dependencies among temperature, humidity, radiation, and precipitation, improving robustness under partial missingness.
    \item \textbf{Vegetation indices}: K-nearest neighbours (KNN) imputation is used with district--season contextual grouping, ensuring that gap-filling respects local agro-climatic conditions and phenological patterns.
    \item \textbf{Soil and topographic variables}: median imputation is employed, reflecting their quasi-static nature and reducing sensitivity to outliers.
    \item \textbf{Outlier handling}: extreme values are winsorised at the 1\% level to limit the influence of measurement noise and retrieval artefacts.
    \item \textbf{Feature scaling}: robust scaling based on median and interquartile range is applied to all numeric features, ensuring comparability across heterogeneous data sources while maintaining resistance to heavy-tailed distributions.
\end{itemize}

\subsubsection{Baseline Models}

To validate the quality and representativeness of the datasets generated by UniCrop, a set of widely used and well-understood machine-learning models is employed. These models are not introduced as methodological contributions, but rather as \emph{reference baselines} commonly reported in the crop yield prediction literature.

The following models are trained:

\begin{itemize}
    \item \textbf{LightGBM}: gradient-boosted decision trees suitable for heterogeneous, non-linear feature spaces.
    \item \textbf{Random Forest}: an ensemble of bagged decision trees providing robust performance with minimal tuning.
    \item \textbf{Support Vector Regression (RBF kernel)}: a kernel-based method capturing non-linear relationships.
    \item \textbf{ElasticNet}: $\ell_1$--$\ell_2$ regularised linear regression serving as a transparent linear baseline.
\end{itemize}

Hyperparameter tuning is intentionally limited to reasonable default ranges to maintain methodological simplicity and reproducibility. The objective of this step is not to maximise predictive performance, but to demonstrate that the features constructed by UniCrop are sufficiently informative to support competitive baseline modelling. Users are encouraged to apply more advanced models or extensive tuning strategies according to their specific research objectives, as improved performance is likely achievable beyond the scope of this study.

\subsubsection{Ensemble Modelling}

To assess whether complementary strengths of individual baseline models can be exploited, a simple ensemble is constructed using out-of-fold predictions. For each training fold, predictions from all base learners are retained and combined only at the validation stage, ensuring that ensemble optimisation does not introduce information leakage.

The ensemble weights are estimated by solving a constrained least-squares optimisation problem:

\[
\min_{\mathbf{w}} \left\| y - \sum_{i=1}^{4} w_i \hat{y}_i \right\|^2
\quad \text{s.t.} \quad
w_i \ge 0,\;
\sum_{i} w_i = 1,
\]
where $\hat{y}_i$ denotes the out-of-fold predictions from the $i$-th base model. Non-negativity and sum-to-one constraints ensure interpretability and prevent over-reliance on any single model. Optimisation is performed using the Sequential Least Squares Programming (SLSQP) algorithm.

The ensemble is not intended to maximise performance, but to provide a stable and interpretable aggregation of baseline predictors, further illustrating the consistency and usefulness of the UniCrop-generated feature set.

\subsubsection{Evaluation Metrics}

Model performance is assessed using multiple complementary metrics to capture both absolute and relative prediction errors:

\[
\mathrm{RMSE}, \quad \mathrm{MAE}, \quad R^2, \quad \mathrm{MAPE}.
\]

Root Mean Squared Error (RMSE) emphasises large deviations and is sensitive to extreme errors, while Mean Absolute Error (MAE) provides a more robust measure of average prediction accuracy. The coefficient of determination ($R^2$) quantifies the proportion of yield variance explained by the model, and Mean Absolute Percentage Error (MAPE) offers an intuitive relative error measure suitable for agronomic interpretation.

In addition to predictive accuracy, model interpretability is evaluated using SHAP (Shapley Additive Explanations). SHAP values are computed for the strongest-performing baseline model to analyse both global feature importance and local, instance-level contributions, enabling assessment of whether predictions rely on agronomically meaningful signals across data sources.

\section{Case Study: Validation of the UniCrop Pipeline Using Rice Yield Data}

To evaluate the effectiveness, robustness, and generalisability of UniCrop, a validation study was conducted using a rice yield dataset comprising 557 field-level observations. This case study demonstrates how harmonised, multi-source environmental data generated by UniCrop can be utilised to support accurate and interpretable yield predictions using standard machine-learning models. Detailed information on the dataset is available through the original challenge documentation (\href{https://challenge.ey.com/challenges/past/level-1-crop-identification-global?id=637e2d535712cf0015c7691f}{link}) and the accompanying academic publication by Yewle et al.~\citep{Yewle2025}. 

The objective of this analysis is not to maximise predictive accuracy, but rather to demonstrate that UniCrop reliably produces high-quality, analysis-ready datasets that enable robust baseline modelling without manual data engineering.

\subsection{Study Area and Spatial Context}

\begin{figure}[h!]
\centering
\includegraphics[width=\linewidth]{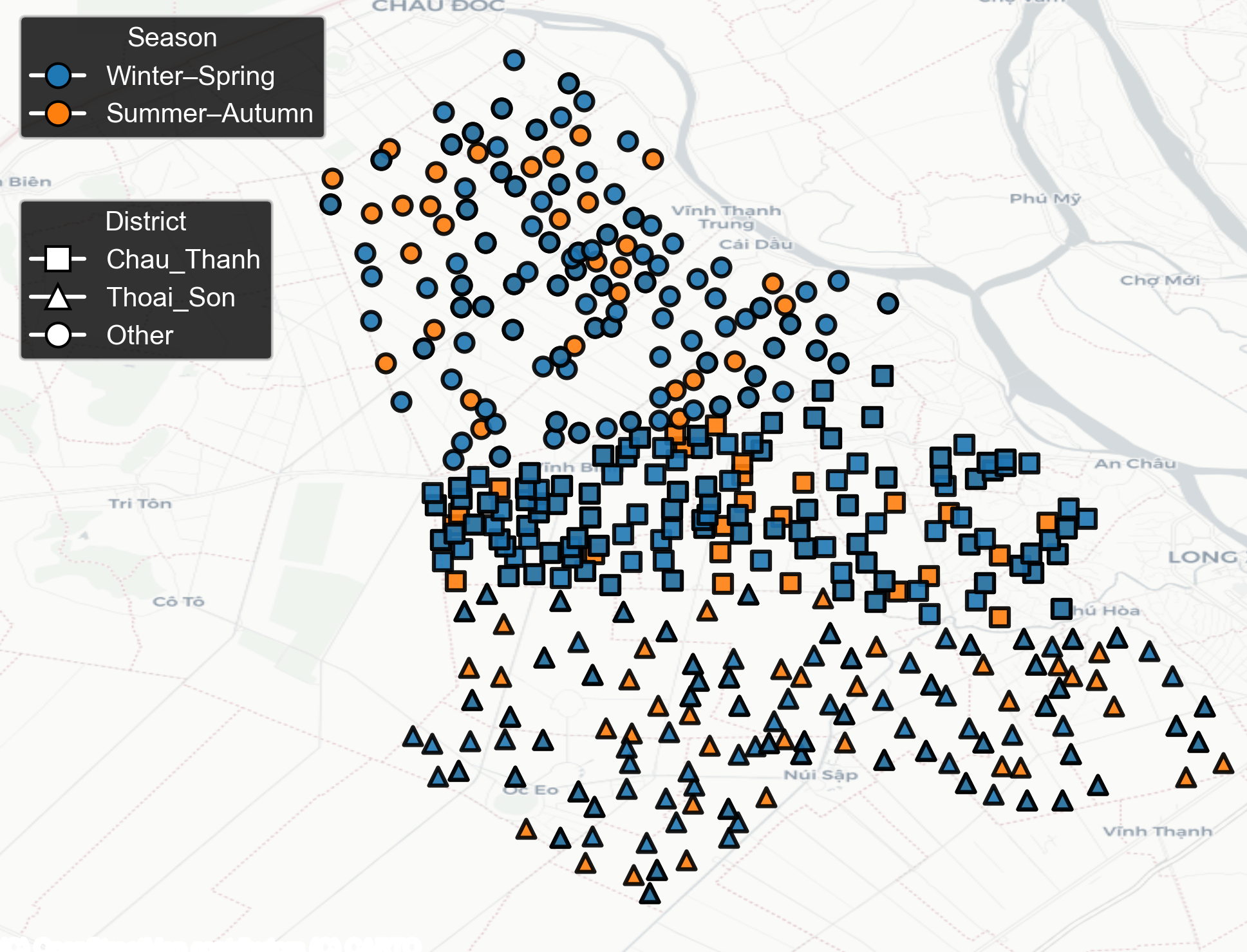}
\caption{Spatial distribution of field parcels included in the case study, colour-coded by growing season.}
\label{fig:spatialmap}
\end{figure}

The dataset spans multiple agricultural districts representing distinct climatic and soil conditions. To provide spatial context, Figure~\ref{fig:spatialmap} shows the distribution of the 557 field parcels, colour-coded by season. Spatial heterogeneity in field locations underscores the importance of harmonising multi-source environmental predictors.

\subsection{Dataset Description}

The reference dataset includes geolocated field parcels with recorded yield measurements (kg/ha), administrative attributes (district, season), and harvest dates. These serve as the seed inputs for UniCrop; all environmental variables are obtained exclusively through automated pipeline execution. After running the full UniCrop process, the resulting master dataset contained approximately 160 candidate features spanning meteorology, vegetation indices, radar backscatter, soil composition, and topography.

\subsection{Exploratory Data Assessment}

The distribution of rice yield values exhibited a unimodal shape centred around approximately 6{,}600\,kg/ha, with mild right skewness. Extended analysis shows that the differences across growing seasons (Summer--Autumn vs.\ Winter--Spring) reflect known climatic and spatial heterogeneity in the region.

\begin{figure}[h!]
    \centering
    \includegraphics[width=\linewidth]{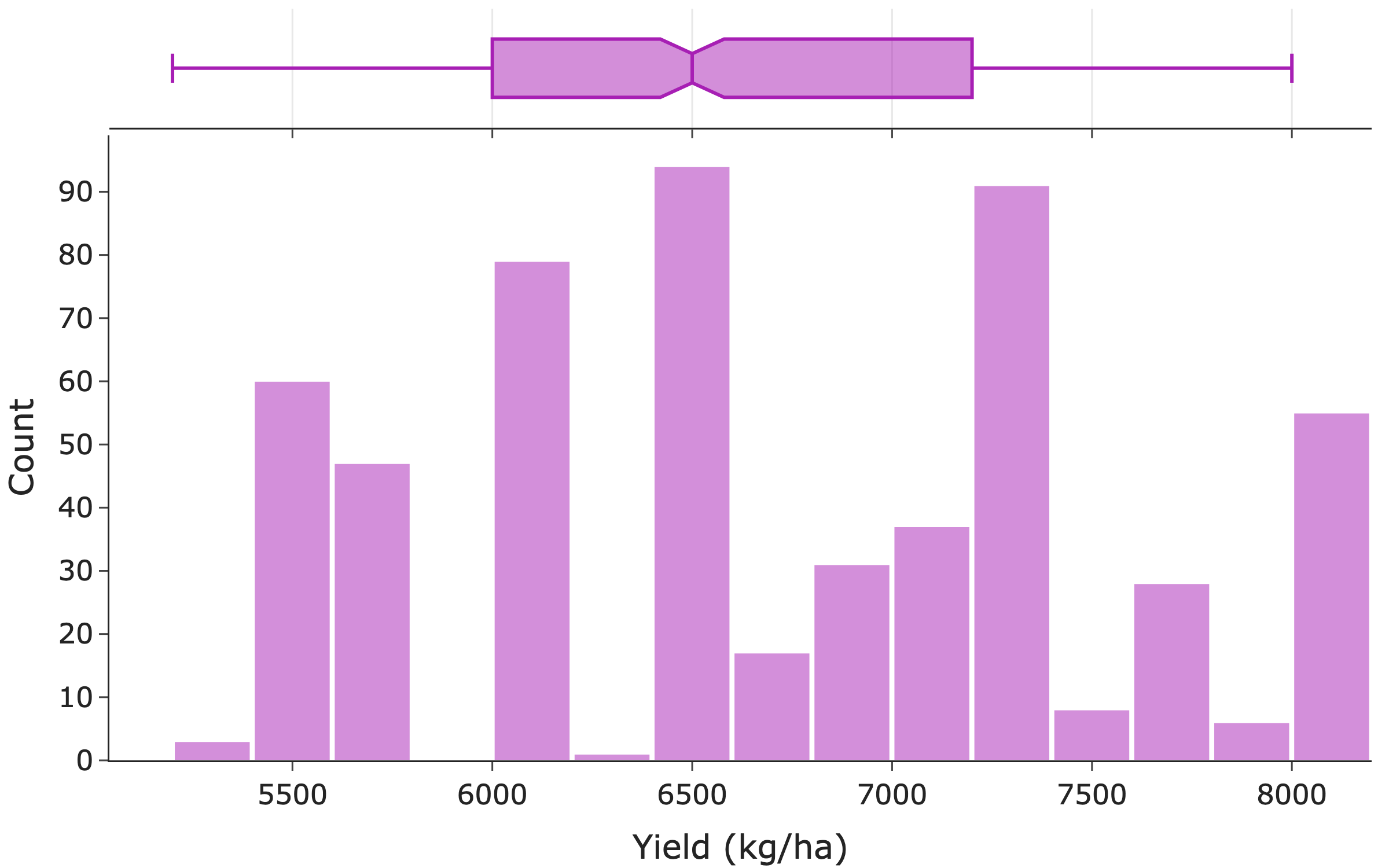}
    \caption{Distribution of rice yield (kg/ha).}
    \label{fig:yielddist}
\end{figure}

Vegetation-related features exhibited higher missingness during the Summer--Autumn season due to monsoon-related cloud cover, while meteorological and soil properties were largely complete. This informed the family-specific imputation strategy applied during cross-validation.

\begin{figure}[h!]
    \centering
        \includegraphics[width=\linewidth]{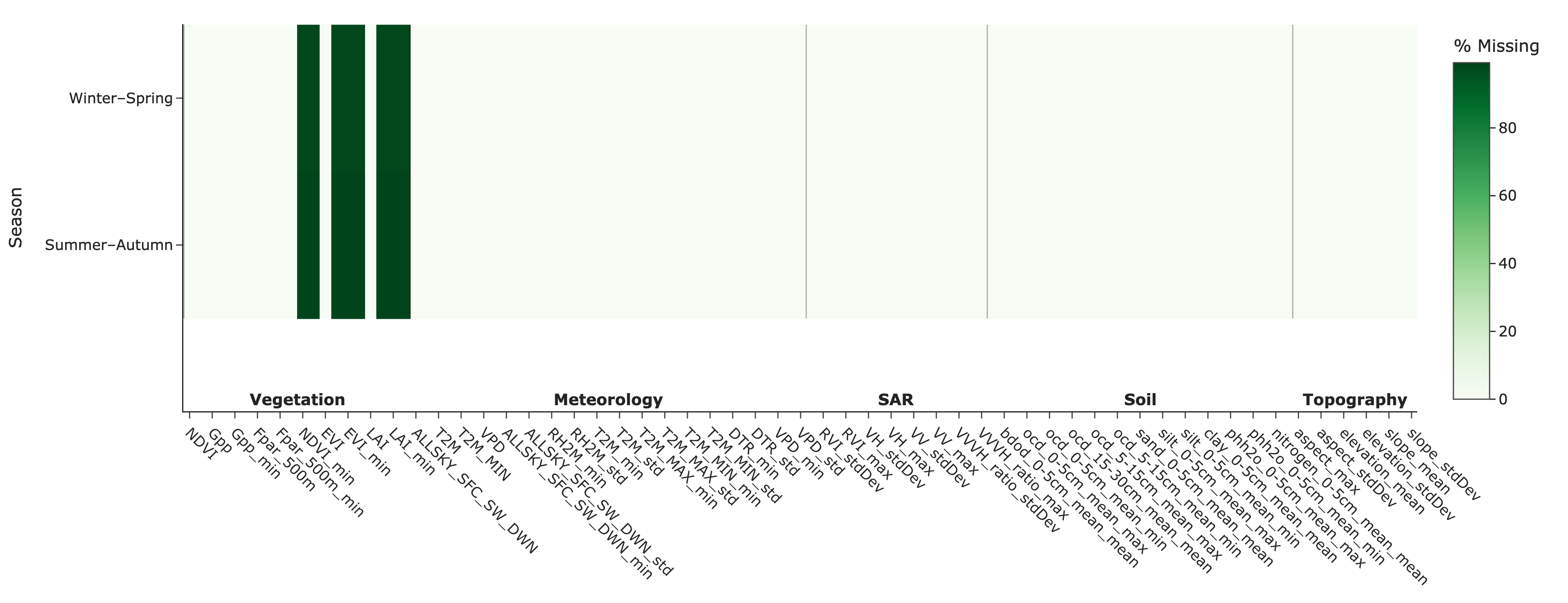}
    \caption{Seasonal missingness patterns for vegetation indices.}
    \label{fig:missingness}
\end{figure}

\subsection{Feature Reduction and mRMR Selection}

Following schema alignment, UniCrop applies a multi-stage feature reduction strategy designed to remove redundancy, enhance interpretability, and ensure that all major environmental data families remain represented. The initial filtering step eliminates near-zero variance features and prunes highly collinear variables ($r \ge 0.98$), retaining only those with the strongest bivariate association with yield based on mutual information. Agronomic feature engineering further enriches the dataset by incorporating thermal time, vegetation dynamics, SAR variability metrics, and soil--climate interaction terms.

To obtain a compact yet informative set of predictors, we employ the minimum redundancy maximum relevance (mRMR) algorithm. mRMR ranks features by balancing two criteria: (\emph{i}) high relevance to the response variable, quantified using a combination of mutual information and rank-based correlations; and (\emph{ii}) low redundancy with respect to features already selected. This ensures that the final subset retains diverse information content rather than concentrating on a single data family or feature type. The algorithm is run independently within each training fold to avoid information leakage.

\begin{figure}[h!]
    \centering
    \includegraphics[width=\linewidth]{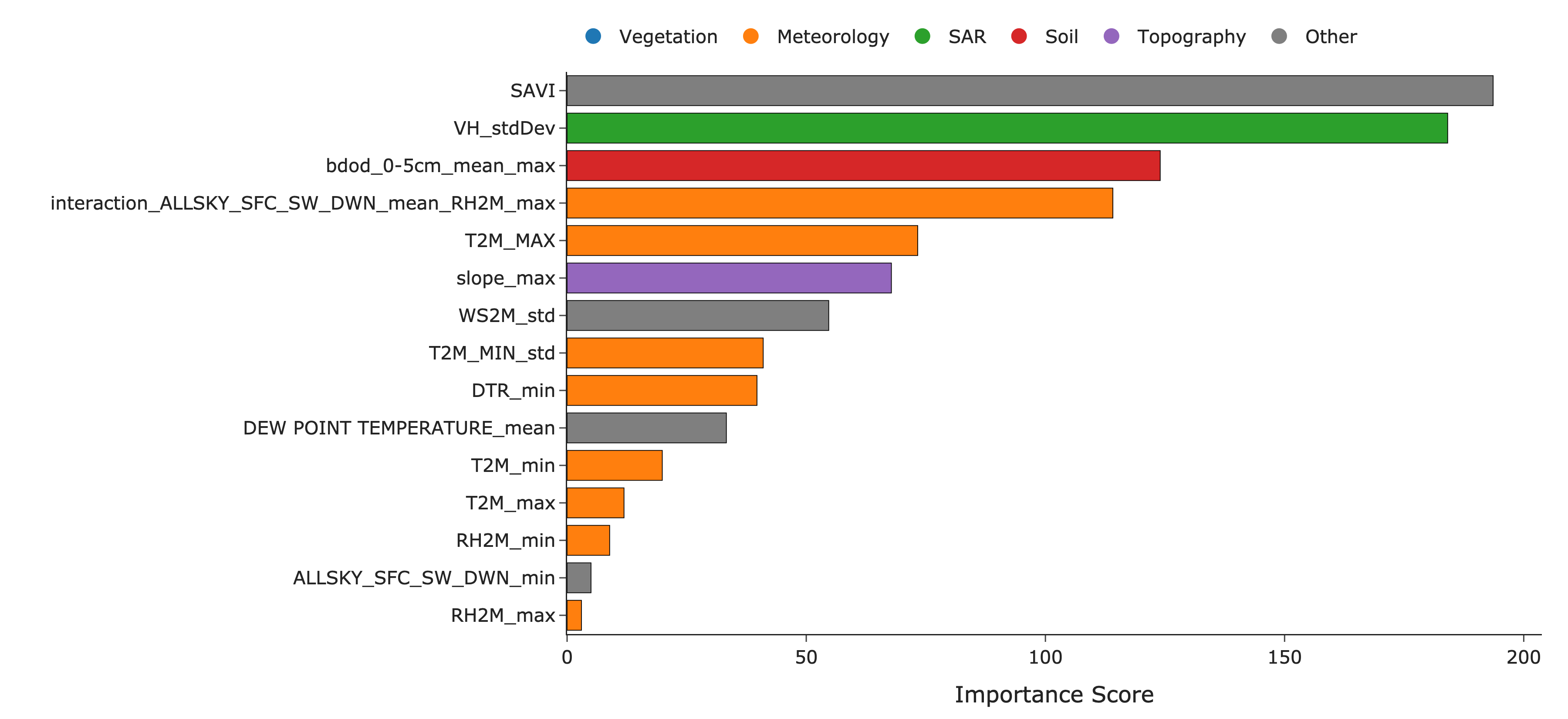}
    \caption{Top 15 predictors selected by the mRMR process. Bars are colour-coded by variable family (meteorology, vegetation, SAR, soil, topography).}
    \label{fig:features}
\end{figure}

Figure~\ref{fig:features} visualises the top 15 selected predictors. Bars are colour-coded by feature family to illustrate the breadth of environmental signals captured by the subset, including vegetation indices, SAR backscatter statistics, meteorological indicators, soil properties, and topographic descriptors. This family-level diversity reflects the multi-source nature of UniCrop’s feature space and demonstrates that the selection process draws from all major data domains rather than overfitting to a single modality. 

The resulting 15-feature set achieves a balance between interpretability and predictive strength. By reducing the original $\sim$160 candidate variables to a well-structured and diverse subset, mRMR provides a principled foundation for downstream modelling, while maintaining representation across all major environmental drivers of rice yield. The presence of vegetation, meteorological, SAR, soil, and topographic variables among the selected predictors highlights UniCrop’s ability to preserve complementary information from heterogeneous data sources.

\subsection{Model Performance}

Four baseline models, LightGBM, Random Forest, Support Vector Regression, and ElasticNet, were trained under a strict 5-fold cross-validation framework, where all preprocessing and feature selection were carried out independently within each fold. This ensures honest evaluation without information leakage.

Table~\ref{tab:performance} summarises the out-of-fold performance metrics. LightGBM achieved the lowest RMSE and highest $R^{2}$ among single models, while a constrained linear ensemble of all base learners provided a small but consistent improvement.

\begin{table}[h!]
\centering
\caption{Cross-validated performance metrics for baseline and ensemble models.}
\label{tab:performance}
\begin{tabular}{lcccc}
\toprule
\textbf{Model} & \textbf{RMSE (kg/ha)} & \textbf{MAE (kg/ha)} & \textbf{$R^{2}$} & \textbf{MAPE (\%)} \\
\midrule
LightGBM & 465.1 & 378.6 & 0.6576 & 5.72 \\
Random Forest & 480.0 & 392.7 & 0.6353 & 5.92 \\
SVR (RBF) & 526.0 & 425.9 & 0.5621 & 6.43 \\
ElasticNet & 467.6 & 380.2 & 0.6539 & 5.73 \\
\textbf{Ensemble} & \textbf{463.2} & \textbf{375.0} & \textbf{0.6604} & \textbf{5.66} \\
\bottomrule
\end{tabular}
\end{table}

\begin{figure}[h!]
    \centering
       \includegraphics[width=\linewidth]{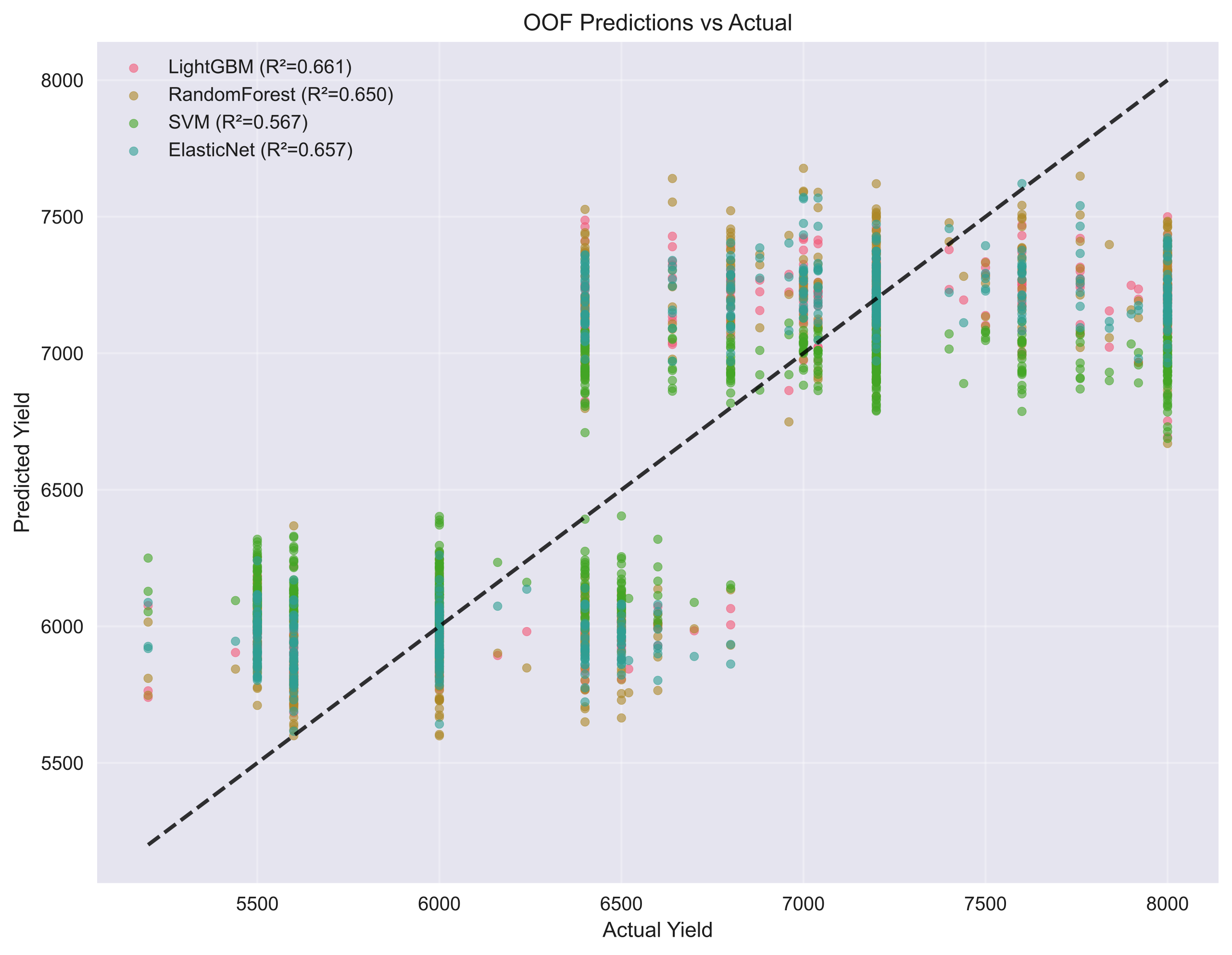}
    \caption{Out-of-fold predictions vs.\ observed yields.}
    \label{fig:scatter}
\end{figure}

Figure~\ref{fig:scatter} visualises the relationship between predictions and observations. Predictions from LightGBM and ElasticNet aligned most closely with the identity line, indicating lower error variance.

\begin{figure}[h!]
\centering
       \includegraphics[width=\linewidth]{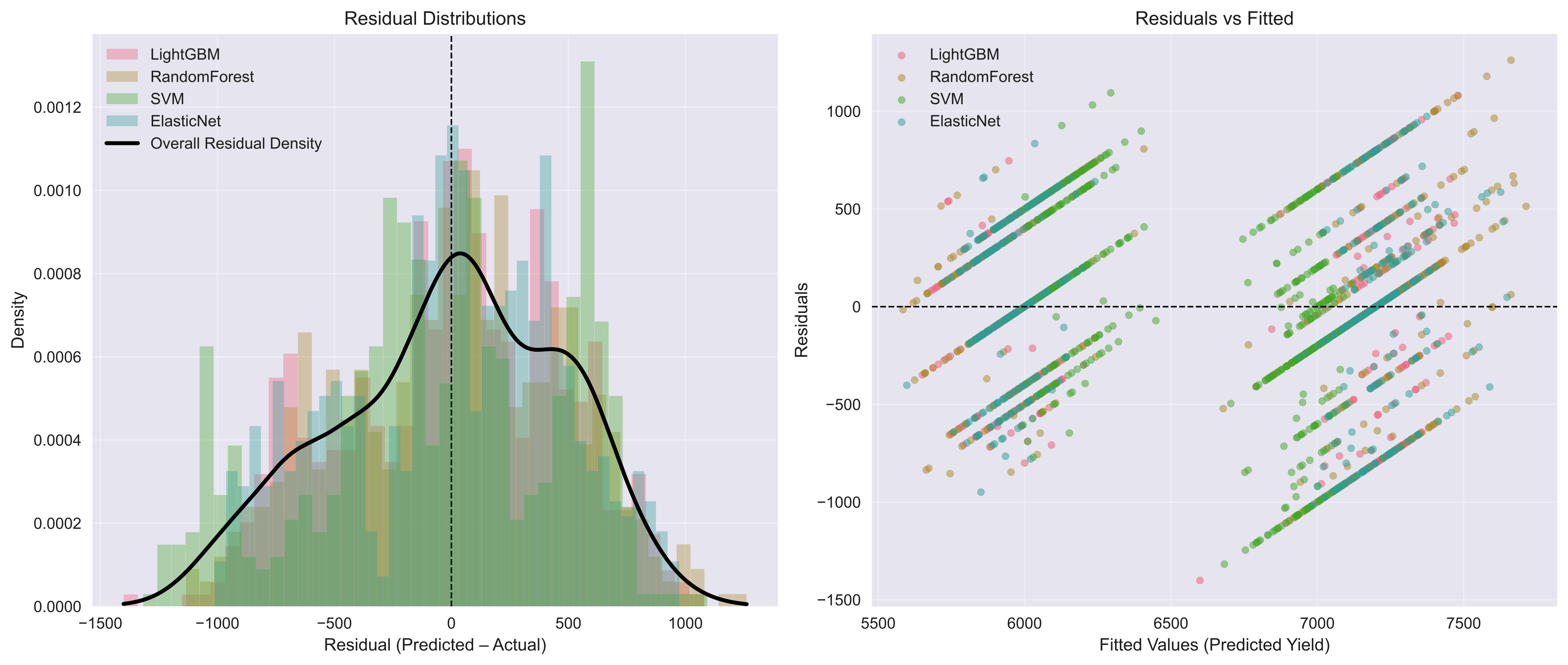}
\caption{Residual distributions and residuals vs.\ fitted values for the LightGBM model.}
\label{fig:residuals}
\end{figure}

To further examine model behaviour, Figure~\ref{fig:residuals} presents a residual distribution plot and residuals versus fitted values. Residuals were approximately centred around zero, with no major heteroskedastic patterns, suggesting that the engineered and selected features provide stable model behaviour.

\subsection{Interpretability via SHAP Analysis}

Global SHAP (Shapley Additive Explanations) values were computed for the LightGBM model to understand feature influence. Several patterns were consistent with agronomic knowledge:

\begin{itemize}
    \item high maximum relative humidity and stable night-time temperatures were associated with higher yield,
    \item high temperature variability and extreme heat had negative impacts,
    \item SAR texture metrics highlighted structural differences in the canopy linked to lodging or uneven growth,
    \item solar radiation interacted non-linearly with diurnal temperature range.
\end{itemize}

\begin{figure}[h!]
\centering
           \includegraphics[width=\linewidth]{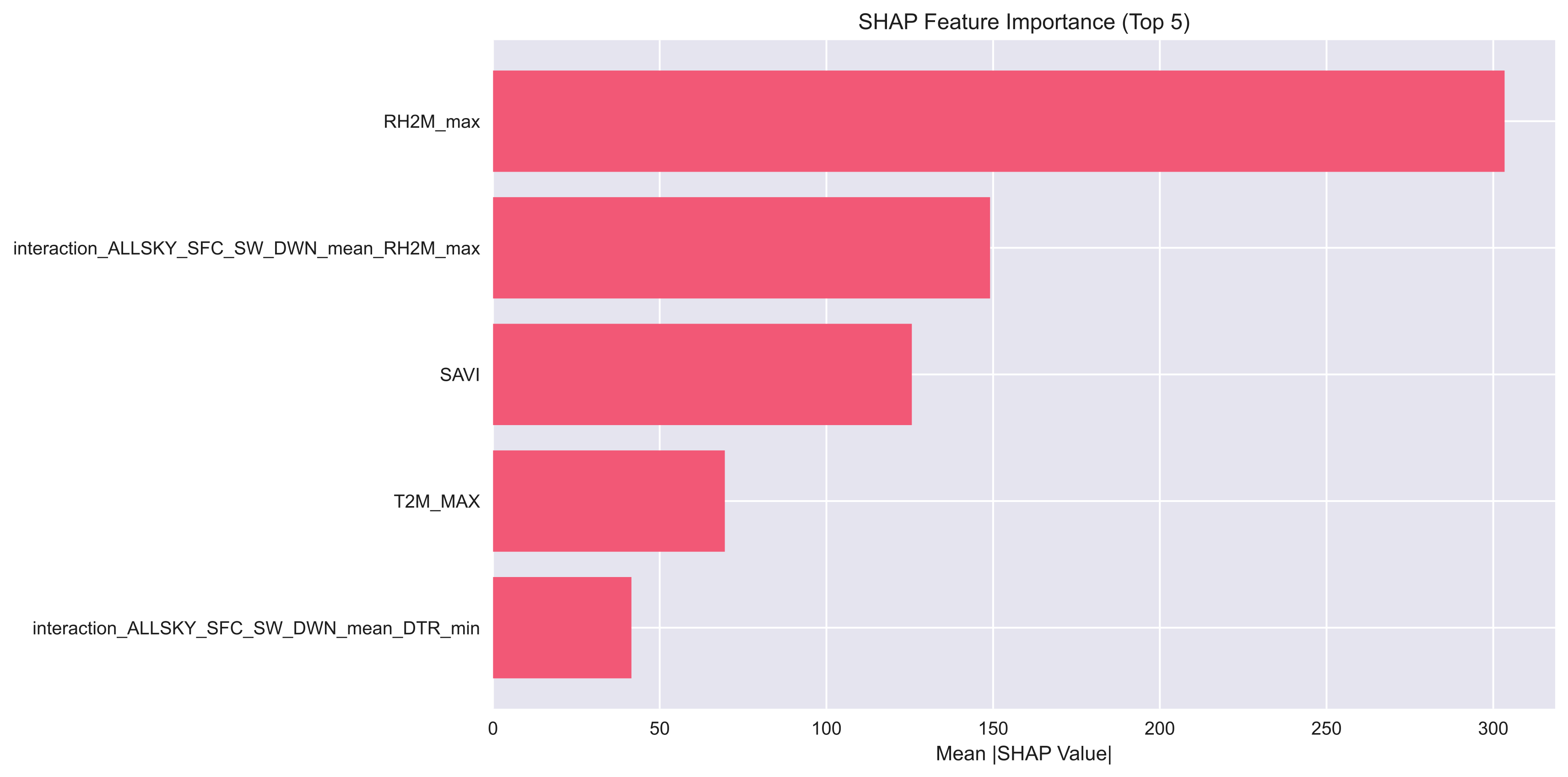}
    \caption{Global SHAP Importance plot for the Top 5 features.}
    \label{fig:shap}
\end{figure}

These interpretability results confirm that the model relies on agronomically meaningful signals across meteorology, vegetation, SAR, soil, and topography, reflecting the multi-modal nature of the UniCrop dataset. These patterns are consistent with established agronomic understanding of rice growth, indicating that UniCrop-derived predictors capture meaningful environmental drivers rather than spurious correlations.

\subsection{Summary of Validation Findings}

The case study demonstrates four key aspects of UniCrop’s effectiveness:

\begin{enumerate}
    \item \textbf{High-quality, harmonised predictors}: The pipeline reliably integrates diverse environmental datasets into a compact, informative predictor set.
    \item \textbf{Strong baseline predictive performance}: Even without model-specific tuning, the processed dataset supports competitive performance among classic ML models.
    \item \textbf{Stable and interpretable residual structure}: Residual diagnostics show well-behaved model error distributions.
    \item \textbf{Robust generalisation}: Performance remains stable under seasonal and spatial shifts.
\end{enumerate}

Taken together, these findings validate UniCrop as a robust, universal data pipeline suitable for scalable yield modelling across different regions and crops.

\section{Discussion and Limitations}

The UniCrop pipeline provides a unified, scalable, and data-centric foundation for crop yield prediction. By automating multi-source data acquisition and harmonisation, the framework addresses a persistent bottleneck in agricultural analytics: the need for bespoke, time-consuming pipelines tailored to specific studies. The case study on rice yield prediction demonstrates that UniCrop can produce high-quality, analysis-ready datasets capable of supporting competitive predictive performance using only standard machine-learning models. This section synthesises the broader implications of UniCrop, its practical value for agricultural modelling, and the remaining limitations that motivate future work.

\subsection{Contributions and Practical Implications}

A key contribution of UniCrop is the decoupling of data specification from implementation. Through a flexible configuration file, users can modify crop type, region, or temporal coverage without altering the underlying codebase. This design enables UniCrop to function as a reusable tool rather than a single-use pipeline, facilitating reproducibility and lowering technical barriers for researchers and practitioners. As agricultural modelling becomes increasingly reliant on multi-modal environmental data, such standardisation offers a pragmatic route toward operational deployment.

The pipeline’s multi-source integration, spanning Sentinel-1/2 imagery, MODIS vegetation dynamics, reanalysis climate records, agro-climatological variables, soil properties, and terrain information, ensures that the resulting dataset captures diverse environmental drivers. The mRMR-based reduction to a compact set of 15 predictors demonstrates that much of the predictive power can be distilled into a small, interpretable subset while still leveraging multiple data families. The robustness of the case study results, including the consistency of SHAP interpretations with agronomic knowledge, emphasises the importance of well-structured data engineering in supporting model reliability.

From an applied perspective, UniCrop can support workflows such as regional monitoring systems, early warning tools, or large-scale benchmarking studies. The ability to generate consistent feature sets across different crops and geographies provides a basis for comparative analysis and transferable modelling frameworks.

\subsection{Limitations and Operational Challenges}

Despite its strengths, UniCrop exhibits several limitations that require consideration. First, the claim of crop-independence has been validated using only a single crop (rice) in a specific region. Although the pipeline is technically generalisable, empirical validation across diverse agro-climatic contexts, including temperate cereals, horticultural crops, and drought-prone systems, is necessary to confirm universality.

Second, the accuracy of UniCrop outputs is inherently constrained by the quality and availability of public data sources. Optical imagery remains susceptible to persistent cloud cover, which can reduce the reliability of vegetation indices in tropical or monsoon-dominated regions. While the inclusion of SAR data mitigates this issue, SAR-derived features capture different aspects of crop structure and do not fully substitute for optical metrics.

Third, UniCrop does not currently include management variables such as irrigation schedules, fertiliser use, planting density, or cultivar information. These factors can substantially influence yield variation, particularly at the field scale. Without such management data, model performance may plateau, especially for fine-grained predictions.

Fourth, although UniCrop performs extensive preprocessing, the pipeline presently produces a single ``snapshot’’ of features per field-date pair, rather than incorporating full temporal trajectories. As interest grows in spatio-temporal deep learning, future versions of UniCrop may require extensions to extract sequential, multi-date time series from satellite and climate sources.

Finally, the computational cost of large-scale GEE queries, especially when covering multiple seasons or high-resolution Sentinel products, may pose practical constraints for countries or institutions with limited cloud resources.

\subsection{Future Directions}

Several avenues for future development arise from these limitations:

\begin{itemize}
    \item \textbf{Multi-crop and multi-region validation}: Applying UniCrop to wheat, maize, soybean, and other major crops across diverse climates will empirically test universality.
    \item \textbf{Integration of management data}: Incorporating farmer-reported or farm-management system inputs could greatly enhance fine-scale modelling.
    \item \textbf{Spatio-temporal extensions}: Adding modules for extracting time series (e.g., multi-date NDVI/SAR/ERA5) would support deep sequence models such as LSTMs, TCNs, and transformer-based architectures.
    \item \textbf{SAR–optical fusion}: Cloud-robust composites that combine Sentinel-1 and Sentinel-2 could improve vegetation monitoring in cloud-prone environments.
    \item \textbf{Local high-resolution refinement}: UniCrop could be adapted into a hybrid framework combining global sources with user-supplied UAV imagery or on-farm sensor data.
    \item \textbf{Scalable deployment}: Containerised or cloud-native implementations (e.g., using Kubernetes or serverless functions) may improve performance for national-scale applications.
\end{itemize}

Overall, UniCrop represents a significant step toward reproducible and transferable agricultural analytics. By addressing the data-engineering bottleneck at the core of yield prediction workflows, the framework lays the foundation for robust, scalable, and interpretable modelling across a wide range of agricultural systems.


\section{Conclusion}

This study introduced \textbf{UniCrop}, a universal and configuration-driven data pipeline designed to automate the preparation of multi-source environmental data for crop yield prediction. By integrating satellite imagery, climate reanalysis, agro-climatological variables, soil composition, and topographic information into a unified and harmonised structure, UniCrop directly addresses one of the most persistent challenges in agricultural analytics: the construction of reproducible, scalable, and generalisable data-engineering workflows. The pipeline’s design separates the specification of required features from implementation, offering a modular and extensible architecture that can be rapidly adapted to new crops, regions, and temporal windows.

The validation case study using 557 rice field observations demonstrates that UniCrop generates analysis-ready datasets capable of supporting strong predictive performance from standard machine-learning models. The compact set of 15 predictors selected through mRMR balances interpretability and predictive power while maintaining representation across all major data families. The resulting models, including a constrained ensemble, achieved competitive accuracy without the need for crop-specific tuning, underscoring the importance of a high-quality data pipeline over model complexity. SHAP-based interpretability further confirmed that the models rely on agronomically meaningful relationships, reinforcing the credibility and scientific soundness of the UniCrop-derived feature set.

While the pipeline exhibits strong generality and robustness, several limitations remain. The absence of management data, reliance on public EO sources, and lack of temporal-sequence extraction constrain its applicability in some contexts. Empirical validation across additional crops and agro-climatic regions will also be important to fully establish universality. Nonetheless, these limitations highlight natural directions for future work, including the integration of high-resolution local data, SAR–optical fusion, and support for spatio-temporal deep learning architectures.

Overall, UniCrop represents a significant contribution to data-centric agricultural machine learning. By automating the most labour-intensive stages of data preparation, it enables researchers and practitioners to focus on model innovation, scenario analysis, and decision support. As global agriculture faces intensifying climatic and economic pressures, tools like UniCrop provide a practical and scalable foundation for transparent, transferable, and impactful yield prediction systems.

\bibliographystyle{elsarticle-num}
\bibliography{paper}


\appendix

\section{Feature Mapping Resources and Documentation}

The UniCrop pipeline uses a configuration-driven feature mapping file that specifies the environmental variables to be retrieved from each data source. This mapping defines the dataset identifiers, API parameters, temporal resolutions, and any required derivations (e.g., vegetation indices, radiation interactions). The mapping file enables users to extend UniCrop to new crops or regions by modifying configuration entries rather than pipeline code.

To ensure correctness and reproducibility, the mapping was constructed using the official technical documentation for all data sources integrated into UniCrop. These resources define band names, variable identifiers, quality flags, spatial resolutions, and recommended preprocessing methods. The most relevant documentation sources are listed below.

\subsection{Earth Observation and Remote Sensing Documentation}

\begin{itemize}
    \item \textbf{Sentinel-1 (SAR):} Copernicus Data Space Ecosystem documentation for GRD products, polarisation modes, and radiometric calibration procedures.
    \item \textbf{Sentinel-2 (Optical):} ESA MSI User Guides, band specifications, Scene Classification Layer (SCL), and Level-2A product processing parameters.
    \item \textbf{MODIS Products (MOD13, MOD15, MOD16, MOD17):} NASA LP DAAC Algorithm Theoretical Basis Documents (ATBDs) describing NDVI/EVI, LAI/fPAR, evapotranspiration, and productivity algorithms.
\end{itemize}

\subsection{Climate and Agro-Climatology Documentation}

\begin{itemize}
    \item \textbf{ERA5-Land:} Copernicus Climate Data Store documentation for hourly land-surface variables, aggregation methods, and variable definitions.
    \item \textbf{NASA POWER:} Agroclimatology Data Dictionary, API reference, and variable descriptions for temperature, humidity, radiation, and wind.
\end{itemize}

\subsection{Soil and Topographic Data Documentation}

\begin{itemize}
    \item \textbf{SoilGrids:} ISRIC documentation on soil depth layers, soil property definitions, and machine-learning based spatial predictions.
    \item \textbf{SRTM:} NASA/USGS documentation for elevation data, void-filling, and terrain derivatives (slope, aspect, hillshade).
\end{itemize}

\section{Supplementary Material and GitHub Release}

To support transparency, reproducibility, and reuse, the UniCrop software framework is released as an open-source repository alongside this study. The GitHub version of UniCrop demonstrates the full pipeline using a \textbf{publicly available maize yield dataset from Spain}, rather than the proprietary rice dataset used in the main case study.

All files required to reproduce the GitHub example are included in the repository. These comprise the sample yield dataset, location-to-coordinate mapping files, configuration scripts, feature-mapping tables, and the full data acquisition and modelling code. The example dataset is derived from the Wageningen University \& Research (WUR) AI sample data repository and is temporally subsampled to include harvest years from 2010 onwards.

The GitHub release includes:
\begin{itemize}
    \item A complete example dataset with latitude--longitude coordinates and annual maize yield values
    \item The feature-mapping file (\texttt{unicrop\_feature\_mapping.csv}) specifying data sources, variables, API parameters, and derivation rules
    \item Configuration files (\texttt{config.py}, \texttt{paths.py}) required to initialise data downloads and modelling
    \item Automated pipelines for data acquisition from NASA POWER, Sentinel-2, MODIS, ERA5, SoilGrids and SRTM
    \item A modular modelling and benchmarking workflow that can be rerun without repeating data downloads
\end{itemize}

This open example is intended as a \textbf{methodological demonstration} of UniCrop rather than a claim of optimal yield prediction performance for maize in Spain. Researchers are encouraged to adapt the configuration files, feature mappings, and data sources to other crops, regions, and temporal settings. The separation between data downloading and modelling allows UniCrop to be readily extended while maintaining reproducibility and clarity of experimental design.

The full UniCrop repository, documentation, and sample data are available at:
\begin{center}
\texttt{[\url{https://github.com/CoDIS-Lab/UniCrop}]}
\end{center}

Please note that the rice yield dataset used in the primary case study of this paper cannot be publicly shared due to data usage restrictions. The dataset is subject to proprietary rights held by Ernst \& Young (E\&Y) and was made available to the authors under specific research and confidentiality agreements. As a result, redistribution of the raw rice yield data or associated parcel-level information is not permitted.

Researchers interested in accessing the rice dataset used in this study should contact Ernst \& Young directly to enquire about data availability and applicable licensing conditions. To ensure transparency and reproducibility of the UniCrop methodology despite these restrictions, we provide a fully open and self-contained GitHub example using a public maize yield dataset, along with all configuration files, feature mappings, and pipeline code required to reproduce the software workflow.

\end{document}